\documentclass{optica-article}

\journal{opticajournal} 

\articletype{Research Article}

\usepackage{graphicx, xcolor}
\usepackage{xspace,color}
\usepackage{upgreek}
\usepackage{bm}
\usepackage{amsmath}
\usepackage{multirow}
\usepackage{lineno}
\usepackage{gensymb}

\usepackage{adjustbox}
\usepackage{booktabs}

\newcommand{\etal}{\textit{et al.\@}\xspace}

\newcommand{\um}{\textmu \text{m}\xspace}
\newcommand{\OCDSl}{$\text{OCDS}_{\textit{l}}$\xspace}
\newcommand{\invivo}{\textit{in vivo}\xspace}
\newcommand{\Invivo}{\textit{In vivo}\xspace}
\newcommand{\invitro}{\textit{in vitro}\xspace}

\newcommand{\enface}{\textit{en face}\xspace}
\newcommand{\Enface}{\textit{En face}\xspace}
\newcommand{\degreeC}{{\degree\@C\xspace}}

\graphicspath{{./Figures}{.}}

\begin{document}

\title{\Invivo dynamic optical coherence tomography of human skin with hardware- and software-based motion correction}

\author{%
	Yu Guo\authormark{1},
	Rion Morishita\authormark{1},
	Ibrahim Abd El-Sadek\authormark{1,2},
	Kohei Yamazaki\authormark{4},
	Shingo Sakai\authormark{4},
	Pradipta Mukherjee\authormark{1,3},
	Yiheng Lim\authormark{1},
	Cunyou Bao\authormark{1},
	Keiichi Sugata\authormark{4},
	Shinya Kasamatsu\authormark{5},
	Hiroyuki Yoshida\authormark{4},
	Shuichi Makita\authormark{1},
	and Yoshiaki Yasuno\authormark{1,*}}
    
\address{%
	\authormark{1}Computational Optics Group, University of Tsukuba, Tsukuba, Ibaraki, Japan\\
	\authormark{2}Department of Physics, Faculty of Science, Damietta University, New Damietta City, Damietta, Egypt\\
	\authormark{3}Centre for Biomedical Engineering, Indian Institute of Technology Delhi, New Delhi, India\\
	\authormark{4}Skin Care Products Research, Kao Corporation, Odawara, Kanagawa, Japan\\
    \authormark{5}Human Health Care Products Research, Kao Corporation, Sumida, Tokyo, Japan}
	\homepage{https://optics.bk.tsukuba.ac.jp/COG/}
	\email{\authormark{*}yoshiaki.yasuno@cog-labs.org}

\begin{abstract*} 
\Invivo application of dynamic optical coherence tomography (DOCT) is hindered by bulk motion of the sample.
We demonstrate DOCT imaging of \invivo human skin by adopting a sample-fixation attachment to suppress bulk motion and a subsequent software motion correction to further reduce the effect of sample motion. 
The performance of the motion-correction method was assessed by DOCT image observation, statistical analysis of the mean DOCT values, and subjective image grading. 
Both the mean DOCT value analysis and subjective grading showed statistically significant improvement of the DOCT image quality. 
In addition, a previously unobserved high DOCT layer was identified though image observation, which may represent the stratum basale with high keratinocyte proliferation.
\end{abstract*}

\section{Introduction}
Optical coherence tomography (OCT) is a non-invasive imaging modality that can provide three-dimensional (3D), high-resolution, and label-free imaging of biological tissues.
OCT has been widely applied to \invivo human imaging in ophthalmology \cite{Huang_OCT_1991, Geitzenauer_Retinal_2010} and cardiology \cite{yonetsu_optical_2013, Vignali_cardiology_2014}.
The imaging depth of OCT, ranging from 1 to 2 mm, enables the effective visualization of both epidermal and dermal structures \cite{Welzel_dermatology_2001}. 
Moreover, OCT allows for the quantitative evaluation of various skin parameters including thickness, reflectivity, and scattering properties \cite{Tsugita_epidermal_thickness_2013, Trojahn_SkinReflectivity_2015, Yamazaki_PSOCT_2020}. 
Unlike biopsies or surgical excisions, its non-invasive and label-free nature enables repeated follow-up assessments while minimizing patient discomfort \cite{Tkaczyk_Non_invasive_2017}.

Although OCT offers several advantages for skin imaging, conventional OCT visualizes only tissue structures, not their physiological and optical functions.
Various extensions of OCT have been demonstrated to overcome this limitation.
For example, polarization-sensitive OCT (PS-OCT) has been introduced to detect polarization properties and enables the evaluation of the tissue micro-architecture at scales smaller than the resolution of standard OCT, such as the organization of collagen and elastin fibers \cite{Sakai_skinPS_2011}.
OCT angiography (OCTA) allows for the visualization and quantification of vasculature and blood flow, and provides critical insights into vascular physiology and pathology \cite{Jia_SplitSpectrum_2012, Wang_Micro_AngioGraphy_2007, Fingler_FlowVisual_2007}.
These extensions has been applied to skin imaging, providing detailed and specific visualization of skin structures and functions \cite{Adams_SkinFibrosis_PSOCT_2020, Deegan_skin_OCTA_2019}.

\begin{figure}
	\centering
	\includegraphics{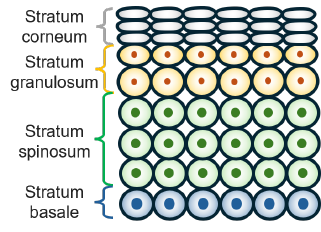}
	\caption{%
		Schematic illustration of four epidermal layers in forearm skin.
		The stratum basale (SB) is located at the base of the epidermis and immediately above the dermal-epidermal junction.
		The keratinocytes proliferate at the SB, making it highly metabolically active.
	}
	\label{fig:Epidemic_layer}
\end{figure}
Although these OCT extensions (i.e., PS-OCT and OCTA) were successful in dermatology, they do not capture intracellular activities and tissue metabolism.
For example, the epidermis includes multiple layers \cite{lenkiewicz_epidermal_2019} as schematically illustrated in Fig.\@ \ref{fig:Epidemic_layer}.
The basal cells in the stratum basale (SB) exhibit high mitotic activity associated with cellular regeneration and turnover \cite{moreci_epidermal_2020}.
This cellular activity may cause intracellular motions.
An imaging modality that can measure the intracellular motions may contrast the SB and characterize its cellular activity.

Dynamic optical coherence tomography (DOCT) is a new modality that noninvasively images intracellular activity \cite{azzollini_dynamic_2023, Ren_DyCOCT_2024}. 
DOCT statistically analyzes the temporal characteristics of time-sequential OCT signals and provides detailed visualization of metabolic activities \cite{Apelian_exvivoDOCT_2016, Monfort_retinal_organoid_2023, Muenter_dynamic_contrast_2020, el-sadek_three_dimensional_2021, morishita_label_free_2023, mukherjee_label-free_2022, AbdElSadek2023, AbdElSadek2024}.

The sensitivity of DOCT to the tissue activity is rooted in long-time-sequence and multiple-frame acquisition at the same location on the sample.
Although the long acquisition time of such a frame sequence is crucial, it poses challenges for \invivo imaging. 
Namely, involuntary motions including subject movement, tissue pulsation, respiration, and other perturbations, such as environmental vibrations and fluctuations, may occur during acquisition, causing inter-frame misalignment of the OCT images.

This misalignment will significantly degrade the image quality of DOCT and imposes substantial difficulty on the practical application of DOCT to \invivo measurements.

Several approaches have been considered to suppress the involuntary motions and/or image misalignment, which can be categorized into two types: hardware-based and software-based approaches.
One example of a hardware-based approach is motion-tracking \cite{mozaffari_retinal_2022}, which has been applied for retinal imaging.
Although this approach effectively suppresses the motion effect, it requires substantial augmentation of the hardware, which results in high complexity and cost of the system.
Another and simpler hardware solution is a sample-fixation attachment \cite{ryabkov_local_2022, Blatter_lesions_2012}. 
Although it requires contact measurement, this approach is low cost, uses simple hardware, and can effectively reduce motion artifacts.

Software-based approach primarily focus on image registration techniques.
The majority of them use cross-correlation or image similarity metrics to align sequential 2D frames \cite{baghaie_involuntary_2017} or volumes \cite{Kurokawa_3DRegestration_2020} and mitigate motion-induced misalignment. 
2D methods are widely used for retinal OCTA imaging \cite{Choi_2D_OCTA_2015, Hwang_2D_Registration_2023} and have recently been applied for functional retinal imaging, such as optoretinography \cite{Schmoll_2D_optoretinography_2010, Suzuki_Optoretinography_2013}.
Although hardware- and software-based approaches are independent, they are not mutually exclusive and can be effectively combined to achieve improved motion suppression.

In this study, we demonstrate \invivo DOCT skin imaging with the combined approach of fixation attachment and registration-based motion correction.
The details of the method are presented first, and its utility and performance for \invivo skin imaging is vastly investigated by conducting an imaging study.
Ten human subjects were involved, with measurements taken from their outer and inner forearm skin. 
The ability of this approach to reduce motion artifacts was quantitatively evaluated by analyzing the mean DOCT values.
In addition, human-grader-based observation study and statistical analysis of the grading scores was conducted to evaluate the ability of our approach to visualize previously unrecognizable skin structure that is possible the SB.

\section{Methods}
\subsection{Motion-suppression method} 
\subsubsection{Hardware-based motion suppression}
\begin{figure}
	\centering
	\includegraphics{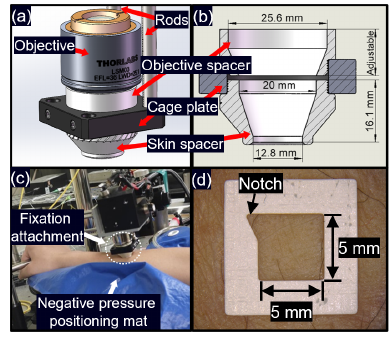}
	\caption{%
            (a) 3D model of the fixation attachment showing the key components, which include the objective, rods, skin spacer, objective spacer, and cage plate. 
            The objective is screwed and fixed to the galvanometer box, which is not shown in this model.
            (b) Cross-sectional schematic illustration of the fixation attachment with dimensional details. 
            (c) Picture demonstrating the posture of the subject for  measurement. 
            The subject's forearm rests on a negative-pressure positioning mat for stabilization, and the skin spacer of the fixation attachment contacts the skin.
            (d) Adhesive marker, which is attached to the skin during the measurement to identify the imaging area (5 mm $\times$ 5 mm).
            The notch serves as a reference point to identify the image orientation.
        }
	\label{fig:Hardware_assemble}
\end{figure}
A sample-fixation attachment was designed to fix the sample during measurements. 
This attachment consisted of a skin spacer that directly contacted the sample (i.e., skin), an optical cage plate (CP33/M, Thorlabs Inc., NJ), and an objective (scan lens) spacer, which was specifically designed for an LSM03 scan lens (Thorlabs), as illustrated in Fig.\@ \ref{fig:Hardware_assemble}(a).
For assembly, the probe was securely attached to the OCT probe unit using rods with 6-mm diameter (ER2, Thorlabs), providing stable and precise alignment.

The skin spacer and objective spacer were both designed using 3D CAD software (SolidWorks, Dassault Syst\`emes, France) and fabricated with a 3D printer (Ender 3 Pro, Creality 3D, China).
The dimensions of these spacers are shown in Fig.\@ \ref{fig:Hardware_assemble}(b). 
The height of the objective spacer is adjustable to ensure that the focus position is set at the depth of interest.
Thermoplastic polyurethane (TPU) was selected as the filament material for its flexibility and durability. 
The 3D-printing process was optimized using slicing software (Cura, Ultimaker, Netherlands).
The nozzle diameter for printing was 0.4 mm, and the nozzle and bed temperatures were set to 215 and 45 \degreeC, respectively. 
The printing speed was configured at 45 mm/s, with a layer height of 0.1 mm.
The CAD files of this spacer are available as open-source hardware design \cite{GitHubMotionCorrection}.

Figure \ref{fig:Hardware_assemble}(c) shows an example of the posture of the subjects for measurement. 
The subject’s forearm was placed on a negative-pressure positioning mat (400 mm $\times$ 400 mm, Navis 7-4576-01, AS ONE, Japan) to provide consistent support during measurement.
The skin space directly contacted and applied pressure to the forearm to minimize bulk motion during the measurement process.

\subsubsection{Software-based motion correction}
\label{sec:softCorrection}
Software-based motion correction \cite{morishita_label_free_2023} was applied after acquiring the OCT images.
This correction was an intensity-image-based 2D correction algorithm consisting of image registration and shifting processes.

The first step of this method was registration of the sequential OCT images, which were obtained at the same location of the sample for DOCT, by calculating the cross-correlation across multiple frames. 
Each frame in the sequence was compared with the center frame, which was the 16th of 32 frames in our specific implementation, using a registration function (skimage.registration.phase\_cross\_correlation(), scikit-image 0.20.0 in Python 3.11.5).
To achieve subpixel precision in shift estimation, the upsampling factor of the registration function was set to 10, allowing for a resolution of 1/10 pixel.

After estimating the shift using the registration process, the shift was corrected by de-shifting each image with a shift function (scipy.ndimage.shift(), SciPy 1.11.4).
This function performed third-order-spline-interpolation-based subpixel shifts on each frame.
It should be noted that both registration and de-shifting were performed on dB-scaled intensity images.
In addition to the hardware-motion suppression, these software steps further minimized the motion artifacts.
The source code of the motion correction software is available under an open source license (MIT license) at a GitHub repository \cite{GitHubMotionCorrection}.

\subsection{OCT system} 
A swept-source Jones-matrix OCT system with an A-line rate of 50 kHz and a center probe wavelength of 1310 nm \cite{li_three-dimensional_2017, Miyazawa_PSOCT_system_2019} was used in our study.
The lateral and axial resolutions were 18 and 14 \um in tissue, respectively, corresponding to pixel separations of 1.95 and 7.24 \um.

Although this OCT system enabled polarization-sensitive imaging, we only used the polarization-insensitive images, which were obtained by averaging the intensities from four polarization channels.

\subsection{Dynamic OCT imaging}
\subsubsection{Scanning protocol}
\label{sec:scanProtocol}
\begin{figure}
    \centering
    \includegraphics{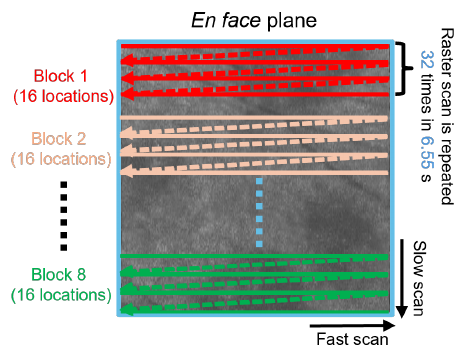}
    \caption{%
        The 32-frame scanning protocol for volumetric DOCT imaging. 
        The \enface plane was divided into eight blocks, with each block containing 16 B-scan locations.
        Each block was repeatedly scanned by a raster-scanning protocol 32 times; hence, 32 OCT frames were acquired at each B-scan location.
    }
    \label{fig:DOCT_scanning}
\end{figure}
For dynamic OCT imaging, a 32-frame repeated raster-scanning protocol \cite{el-sadek_three_dimensional_2021} was used.
In this 3D DOCT scanning protocol, the \enface field was divided into eight blocks, as shown in Fig.\@ \ref{fig:DOCT_scanning}.
Each block, which consisted of 16 B-scan locations, was repeatedly and rapidly raster-scanned 32 times in 6.55 s.
Hence, 32 repeated frames were captured at each B-scan location with an inter-frame interval of 204.8 ms.
A total of 4,096 frames (32 frames/B-scan location) was captured in 52.4 s.
Each frame consisted of 512 A-lines.

In addition to the 3D DOCT scanning protocol, a standard four-frame repeating raster-scanning protocol was applied for OCT angiography (OCTA) imaging.
The inter-frame interval was 12.8 ms, and the 4 frames were captured in 38.4 ms.
Complex-correlation-based OCTA (cmOCTA) \cite{makita_noise-immune_2016} was computed from the 4 frames and compared with the DOCT images obtained from 32 frames.

The field of view for both scanning protocols was set to 6 mm $\times$ 6 mm, and an adhesive marker with a 5 mm $\times$ 5 mm imaging window was attached to the skin to mark the scanning area, as shown in Fig.\@ \ref{fig:Hardware_assemble}(d).
The notch in the upper left served to identify the orientation of the sample in the image.

\subsubsection{DOCT algorithm}
For DOCT imaging, we used the logarithmic intensity variance (LIV) as the DOCT algorithm \cite{el-sadek_optical_2020}. 
LIV is the time variance of the dB-scaled OCT intensity and is defined as
\begin{equation}
    \label{eq:LIV}
    \mathrm{LIV}(x, z) = \left\langle \left[ I_{\mathrm{dB}}(x, z, t_i) - \langle I_{\mathrm{dB}}(x, z, t_i) \rangle_t \right]^2 \right\rangle_t,
\end{equation}
where $I_{\mathrm{dB}}(x, z, t_i)$ is the dB-scaled OCT intensity at the lateral position $x$ and the depth position $z$; $t_i$ is the sampling time point of the $i$-th frame, where $i = 0, 1, 2, \dots, N-1$; and $\langle \qquad \rangle_t$ represents time averaging over all points.

In addition to the DOCT, an averaged intensity OCT image was derived by averaging all dB-scaled OCT frames at the same location.

To further enhance the visualization of tissue dynamics, a pseudo-color LIV image was created by combining the averaged OCT intensity and LIV values.
In this image, the pixel brightness corresponded to the OCT intensity and the pixel hue corresponded to the LIV.

\subsection{Sample and measurement}
\label{Section: sample}
To evaluate the motion-suppression effects of our methods \invivo, we recruited 10 subjects aged 24 to 29 years (mean age: 25.1), including 3 males and 7 females.
Both the outer and inner forearms of each subject were measured.

The measurement process was conducted sequentially, starting with the outer forearm and followed by the inner forearm. 
For each subject, the outer forearm was first measured without the fixation attachment (i.e., solely with the scan lens) using the 32-frame DOCT scanning protocol, followed by the 4-frame cmOCTA scanning protocol.
After completing these measurements, the fixation attachment was attached, and volumetric datasets for DOCT and cmOCTA were acquired.
Upon completion of the outer forearm measurements, the procedure was repeated for the inner forearm following the same steps as for the outer forearm.
For all measurements with and without the fixation attachment, the negative-pressure positioning mat was used to stabilize the arm.

For each subject, images with four configurations of motion suppression and correction were obtained.
The configurations include those with both hardware and software methods (HS), only with hardware suppression (H), only with software correction (S), and with no correction (NC).

The present study adhered to the tenets of the Declaration of Helsinki and was approved by the Institutional Review Boards of University of Tsukuba.

\subsection{Study-1: Quantitative analysis of motion correction through LIV values}
\subsubsection{Image pre-processing for quantitative analysis}
The spacer tip, which was a ring, presses the skin and caused the skin surface to curve.
To analyze the results as a function of the depth from the surface, it is essential to detect the skin surface and flatten the volume with respect to the surface.

We used an open-source neural-network-based segmentation method, Segment Anything Model (SAM) \cite{kirillov_segment_2023}, to detect the surface.
Although we needed a small number of manual prompts, which is the selection of a few pixels to designate the target tissue, SAM can segment the tissue region without additional training.

SAM was applied to the averaged dB-scaled intensity cross-section, and the depth positions of the topmost pixel in the segmented skin region were used to define the skin surface.
For some subjects, hairs on the skin were segmented as part of the skin, and hence the top pixels of the hairs were erroneously recognized as the skin surface.
To correct this, volumes with such errors were manually identified, and morphological operations were applied to remove the segmentation error.
We sequentially applied opening and closing operations.
The structuring element was circular with a diameter of 10 pixels for the opening operation, and the diameter of the closing operation was 15 pixels.
In practice, the circular structuring elements were generated as an elliptical structuring element with the same dimensions for two directions by cv2.getStructuringElement(shape = cv.MORPH\_ELLIPSE) (Python version of OpenCV 4.10.0) and the morphological operations were performed by cv2.morphologyEx().

\subsubsection{\Enface slab averages and region of interests}
To analyze and compare tissue structures and dynamics across different depths, \enface slab averages were generated for slabs of four different depths: 20--220 \um, 220--420 \um, 420--620 \um, and 620--820 \um, which roughly correspond to the epidermis and papillary dermis, upper reticular dermis, middle reticular dermis, and deeper reticular dermis, respectively \cite{SandbyMoeller_SC_thickness_2003, Wang_skin_thickness_2022, Villaret_papillary_dermis_2018}.
A comparison among the three modalities, OCT, cmOCTA, and LIV, was conducted using the slab averages from these four depth ranges.

\begin{figure}
    \centering
    \includegraphics{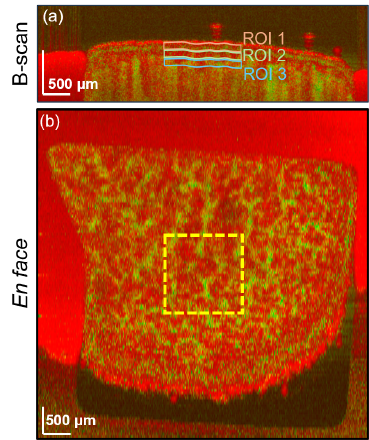}
    \caption{%
        An example of ROI selection.
        (a) Three ROIs were selected at different depths: 20--220 \um for ROI 1 (orange box), 220--420 \um for ROI 2 (green box), and 420--620 \um for ROI 3 (blue).
        The yellow box in (b) illustrates the \enface location of the ROIs.
	}
    \label{fig:ROI_selection}
\end{figure}
For quantitatively analyzing the motion correction effect of our hardware and software method, we selected the center region of the \enface imaging field, as shown in Fig.\@ \ref{fig:ROI_selection}(b).
Three 3D regions of interest (ROIs) were set at each of the first three upper slabs, as shown in Fig.\@ \ref{fig:ROI_selection}(a).

\subsubsection{Statistic analysis of mean LIV values}
\label{sec:ROI_t-test}
The same subjects were measured by two hardware configurations, i.e., with and without hardware motion suppression.
And four types of images are obtained from the two measurement, they are the combinations of with/without hardware motion suppression and with/without software motion correction.
The adhesive marker ensured that the same region of the sample was measured for all configurations.
The average LIV values were calculated for each ROI of each configuration.
To statistically compare the effects of hardware motion suppression and software motion correction, paired t-tests were performed for each configuration at each ROI.
The purpose of this analysis was to determine whether there were significant effects of hardware motion suppression and/or software motion correction.

\subsection{Study-2: Observational evaluation of high-value layers}
\subsubsection{Intra-epidermal high-value layer}
\label{sec:methodIEHV}
The SB is a single-cell layer in the epidermis that is composed of basal keratinocytes \cite{lenkiewicz_epidermal_2019}.
Because this layer is characterized by the cellular proliferation of the basal keratinocytes \cite{moreci_epidermal_2020}, it may exhibit prominent tissue dynamics and high LIV.
As later shown in the Results section (Section \ref{sec:resultIEHV}), we frequently observed a high-LIV layer in the epidermis, which may correspond to the SB.
Since it is not fully confirmed that this high LIV layer was the SB, we do not denote it as SB but denote it as an intra-epidermal high-value (IEHV) layer in the following sections.
In addition, if a high-intensity layer appeared in the OCT image, it will also be denoted as IEHV layer.

\subsubsection{Subjective grading of the IEHV layer}
\label{sec:methGrading}
To assess the visualization quality of the IEHV layer, an observational-grading study was conducted.
The visibility and sharpness of the IEHV were graded by three graders including two skin specialists (Sakai and Yamazaki, denoted as SS and KY, respectively) and one OCT specialist (Yasuno, YY).

The assessment was based on both LIV and OCT intensity images. 
The visibility of the IEHV layer of each image was scored according to the following criteria:
\begin{itemize}
	\item[0:] Invisible.
	\item[1:] Visible but low contrast.
	The IEHV layer is visible but with limited contrast or visible only in some parts of the field.
	\item[2:] Clearly visible.
	The IEHV layer is easily identifiable with high contrast and stands out clearly from the adjacent layers.
\end{itemize}

Similarly, the sharpness of the IEHV layer was scored according to the following criteria:
\begin{itemize}
	\item[0:] Invisible. The IEHV layer is not visible, making it impossible to assess the sharpness.
	\item[1:] Diffusive. The layer appear blurry or diffusive with poorly defined boundaries.
	\item[2:] Sharp. The layer has a clear and well-defined boundary from the surrounding region.
\end{itemize}

The evaluation dataset consisted of 160 B-scan images, which derived from four motion-correction configurations across two scan locations (outer and inner forearm) for 10 subjects, using two image types (OCT and pseudo-color LIV).
One representative B-scan location per volume was used for the grading.
The images were presented to the graders in a randomized order, with the graders blinded to the motion-correction configurations, scan locations, and subjects.

\subsubsection{Statistical analysis for visibility and sharpness scores}
Because the scores were not parametric, a non-parametric test (Wilcoxon signed-rank test) was applied to evaluate differences among motion-suppression/correction configurations.
This analysis supplemented the previous t-test analysis of the quantitative evaluation (Section \ref{sec:ROI_t-test}) and provided further insight into the effects of the motion-suppression/correction methods.
An additional Wilcoxon signed-rank test was conducted to compare the scores of LIV and OCT images with the same configurations for motion-suppression/correction configuration. 
The purpose of this analysis was to highlight the advantages of LIV in visualizing the IEHV layer.

\section{Results}
\subsection{Image observations}
\subsubsection{Observation of the LIV images}
\begin{figure}
    \centering
    \includegraphics{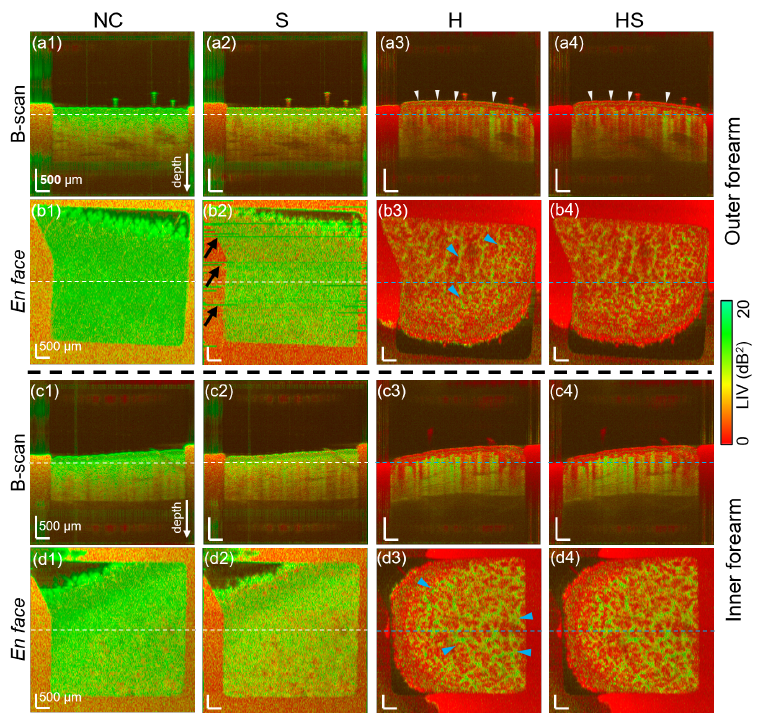}
    \caption{%
        LIV images of the outer and inner forearms.
        The line in the B-scan indicates the position of the corresponding \enface image, vise versa.
        Each column represents different configurations for motion suppression and correction; from the right, with no correction (NC), only with the software method (S), only with the hardware method (H), and with both the hardware and software methods (HS).
        It can be found that the hardware method effectively minimized motion artifacts.
    }
    \label{fig:LIV_images}
\end{figure}
Figure \ref{fig:LIV_images} shows representative B-scan and \enface LIV images of the human outer and inner forearm skin. 
Without hardware suppression and software correction (NC; first column), all LIV images exhibited vast regions with high LIV (green) [Fig.\@ \ref{fig:LIV_images}(a1, b1, c1, d1)].
These regions with high LIV were likely due to motion artifacts.

After applying the software correction but without hardware suppression (S, second column), some regions that originally exhibited high LIV values in the images without correction (NC, first colume) now show lower values (more reddish appearance)[Fig.\@ \ref{fig:LIV_images}(a2, b2, c2, d2)].
However, meaningful structures were still not visualized, and some prominent high-LIV line artifacts were observed [black arrows, Fig.\@ \ref{fig:LIV_images}(b2)].
These artifacts were likely caused by bulk motion exceeding the correction range of the software method.

Even without software correction, the fixation attachment greatly reduced the high-LIV artifacts (H, the thrid column) [Fig.\@ \ref{fig:LIV_images}(a3, b3, c3, d3)]. 
In the \enface image, several high-LIV vessel-like structures became clearly visible [Fig.\@ \ref{fig:LIV_images}(b3, d3), indicated by the light-blue arrowheads].

The software motion correction further suppressed the motion artifacts (HS, the fourth column).
This improvement was particularly evident in the surface region [compare Figs.\@ \ref{fig:LIV_images}(a3) and (a4), white arrowheads], where the very thin artifactual green layer in at the surface was removed by the additional software correction.
These results qualitatively demonstrate the effectiveness of our combined hardware and software motion-correction methods in reducing motion artifacts and improving the visualization of tissue dynamics.

In addition, it is noteworthy that the cross-sectional images of H and HS exhibited distinctive high-LIV layers in the epidermis, which we designated as the IEHV layer, as discussed in Section \ref{sec:methodIEHV}.
The visibility and the sharpness of this layer are extensively investigated in Section \ref{sec:resultIEHV}.

\subsubsection{Descriptive comparison of OCT, OCTA, and LIV}
\label{sec:resultSlabProjection}
\begin{figure}
	\centering
	\includegraphics{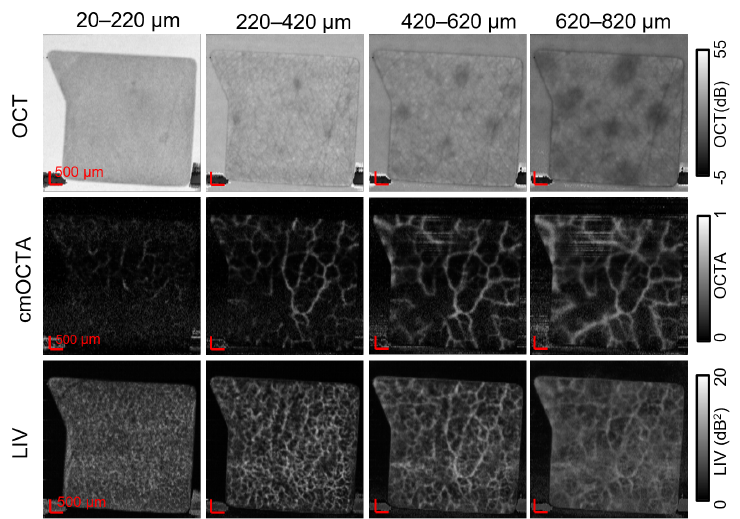}
	\caption{%
		The depth averaged images of outer-forearm skin at four different depth slabs.
		From the top row, each row shows OCT, cmOCTA, and gray-scale LIV images.
		The LIV images show significantly more vessels than the cmOCTA images.
	}
	\label{fig:Vessel_4depth}
\end{figure}
Figure \ref{fig:Vessel_4depth} shows representative \enface slab averages of OCT, cmOCTA, and LIV for four depth ranges.
cmOCTA clearly visualized large vessels.
On the other hand, LIV images exhibit smaller and more intricate vessels that are not as apparent in the cmOCTA images.
This enhanced visualization is attributed to that the LIV is computed from a set of larger number of frames obtained with a longer overall time window (6.55 s) than OCTA (38.4 ms).
This comparison suggests that LIV can be used as very high sensitive OCTA enabling the detection of fine vascular structures that are less apparent in OCTA images.
This point is further discussed in Section \ref{sec:bfImaging}.

\subsection{Quantitative analysis of the mean LIV (Study-1)}
\label{sec:resultMeanLIV}
\begin{figure}
    \centering
    \includegraphics{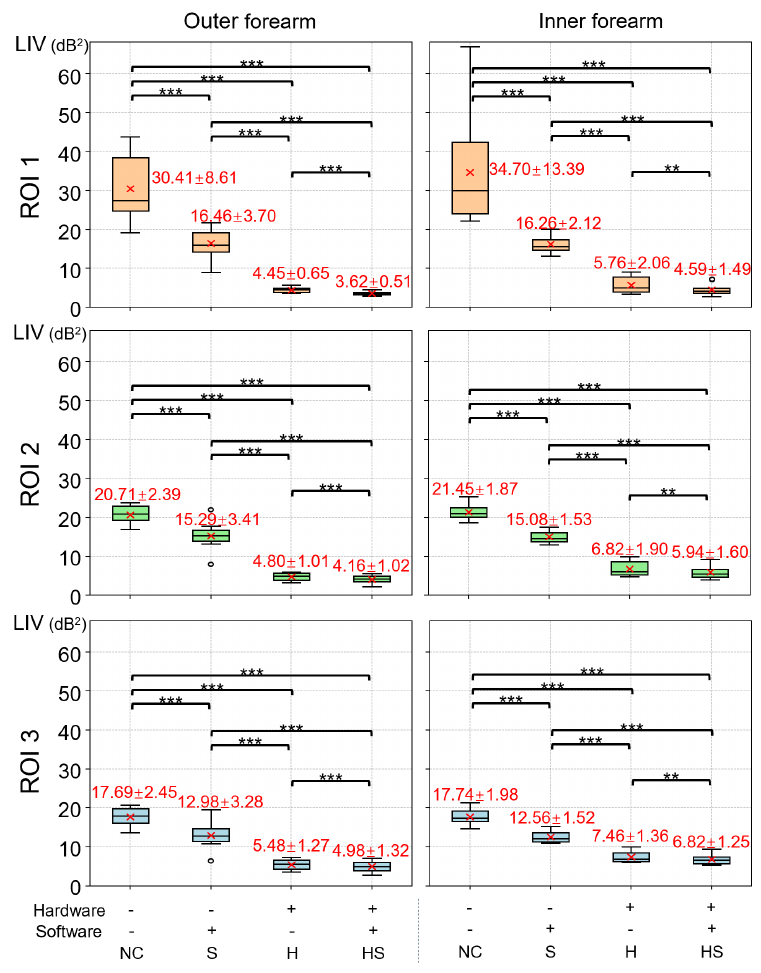}
    \caption{%
        Mean LIVs of 10 subjects at each configuration of motion suppression and correction, measured from the outer and inner forearm.
        The crosses and whiskers indicate the mean, maximum, and minimum over the 10 subjects, and the box and the bar in the box indicate the 25th and 75th percentiles, and the median, respectively.
        Each value in the plots represents the mean $\pm$ standard deviation.
        The stars indicates statistical significance, with ** for $p$ < 0.01 and *** for $p$ < 0.001.
    }
    \label{fig:ROI_box_plot}
\end{figure}
The mean LIV values of each ROI on both sides of the forearm are summarized as box-whisker plots in Fig.\@ \ref{fig:ROI_box_plot}, where each plot compares the four motion-suppression/correction configurations (i.e, NC, H, S, and HS).
The lower and upper bounds of box indicate the 25th and 75th percentiles, and the horizontal line in the box is the median (i.e., 50th percentile).
The crosses represent the mean values, and the whiskers extend to the maximum and minimum values.

The star symbols in the plot indicate a statistically significant difference obtained by a paired t-test, where ** represents $p$-value < 0.01 and *** represents $p$-value < 0.001.
All $p$-values of the t-test are summarized in Tables S1 and S2 in the supplementary material.

The three ROIs and both sides of the forearm exhibited a similar trend of LIV reduction.
The LIV values without fixation attachment (NC and S) were very high, which might be caused by motion artifacts. 
It should be noted that, although the introduction of the software correction reduced the LIV values (i.e., motion artifacts) with statistical significance, they remained very high.

By contrast, for the configurations with fixation attachment (H and HS) there was a statistically significant reduction in LIV values, highlighting the importance of fixation attachment.
Notably, in all cases, HS exhibited statistically significantly lower LIV values than H.
This highlights the additional benefit of software correction when combined with the fixation attachment.

\subsection{Visibility and sharpness of the IEHV layer (Study-2)}
\label{sec:resultIEHV}
\subsubsection{Main results of Study-2}
\begin{figure}
    \centering
    \includegraphics{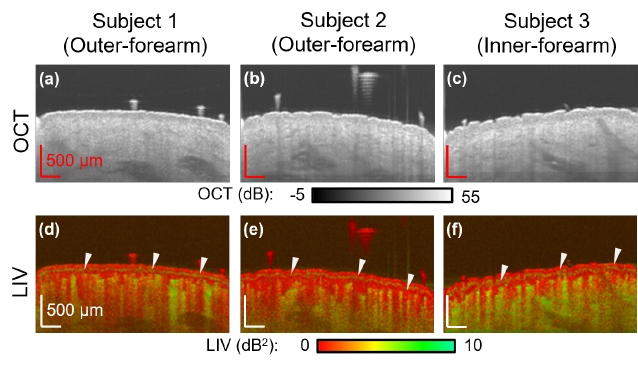}
    \caption{%
        Enlarged cross-sectional OCT and LIV images of the skin surface from three representative subjects.
        The LIV images show thin high-LIV layers (IEHV layers) in the epidermis (arrow heads), which are not visible in the OCT images.
        }
    \label{fig:IEHV_layer}
\end{figure}
Figure \ref{fig:IEHV_layer} shows representative enlarged cross-sectional OCT and LIV images.
In the LIV images, thin high-LIV layer (IEHV layer) are highly distinctive in the epidermal region (white arrows), whereas it remains undetectable in both the OCT and cmOCTA images. 
This highlights the unique capability of LIV to capture dynamic tissue features that are not detectable in either structural OCT or cmOCTA.

\begin{figure}
    \centering
    \includegraphics{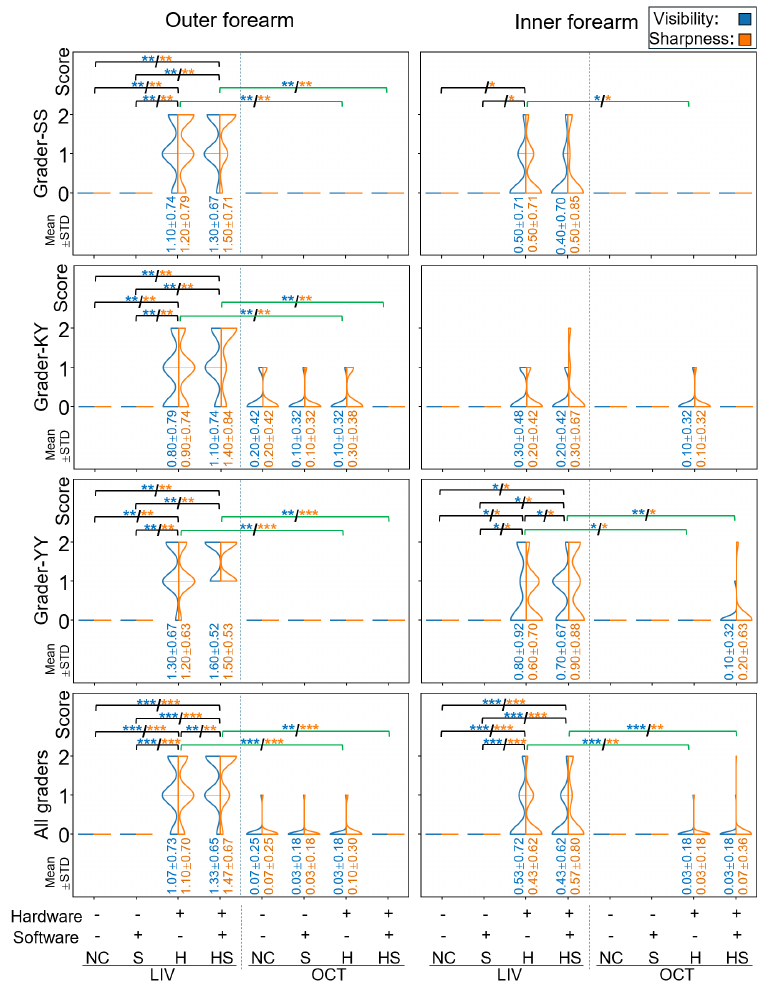}
    \caption{%
        Violin plots displaying the visibility and sharpness scores obtained from three graders (SS, KY, and YY) and their combined assessment (all graders). 
        The blue and orange represent the visibility and sharpness evaluations, respectively. 
        The mean scores and standard deviations are presented under the plots, where the blue and orange correspond to the visibility and sharpness, respectively.
        The rows correspond to the graders and the columns correspond the imaging side of the forearms. 
        Statistical differences between the different motion suppression/correction configurations are indicated by the black lines and stars, and significance between LIV and OCT is highlighted by green lines and stars.
        The $p$-values are summarized in Tables S3 (viability) and S4 (sharpness) in the supplementary material.
    }
    \label{fig:Grader_violin_plot}
\end{figure}
To further highlight the superior capability of LIV for visualizing the IEHV layer compared with OCT, three graders independently scored the visibility and the sharpness of the IEHV layer in the OCT and LIV images.
(The grading criteria can be found in Section \ref{sec:methGrading}.)
Figure \ref{fig:Grader_violin_plot} summarizes the scores of all graders for each type of image, each outer and inner forearm, and each configuration of motion suppression/correction as violin plots.
The blue and orange represent the visibility and sharpness scores, respectively.
All $p$-values of the Wilcoxon signed-rank test are summarized in Tables S3 and S4 in the supplementary material.

It is noteworthy that, in the LIV images, the visibility scores (blue) for configurations without hardware suppression (i.e., NC and S) were all zero.
The application of the hardware suppression (H and HS) notably increased the visibility scores of the LIV images.
On the other hand, the visibility scores for OCT images were zero in almost all cases, regardless of the application of motion suppression/correction.

The statistical significance of the visibility difference among the different configurations of motion suppression/correction were examined using Wilcoxon signed-rank tests.
As expected from the very low visibility of OCT (i.e., zero in almost all cases), the visibility of the IEHV layer in the H and HS LIV images were significantly higher than in the corresponding OCT images, except for the case of inner forearm of Grader 2.
This highlights the superior visualization capability of LIV compared with that of OCT.

Hereafter, we only discuss the LIV images because almost all OCT images had zero visibility.
In the outer forearm, the IEHV layer became significantly more visible with hardware correction for all graders (NC vs H, $p$ < 0.05 for one grader and $p$ < 0.01 for two graders; S vs HS, $p$ < 0.01 for all graders).
By contrast, the additional application of software correction (HS) to the hardware suppression (H) did not result in a significant improvement in visibility for each grader, although there was a significant difference among the combined grading results of the three graders ($p$ < 0.01).
In summary, hardware motion suppression is crucial to making the IEHV layer visible, but software correction can be optional.
However, it might be worth noting that the previously presented mean LIV showed a significant reduction in motion artifacts with software correction (HS vs H and NC vs S, as shown in Section \ref{sec:resultMeanLIV}).
Therefore, it is still worth applying additional software correction.

There was less consistency among graders regarding the visibility of the inner forearm than for the outer forearm.
In addition, the improvements achieved by the hardware and software methods were less significant than those for the inner forearm.
This limited improvement may because of relatively high background LIV values, i.e, residual motion artifacts [see H and HS of ROI 1 in Fig.\@ \ref{fig:ROI_box_plot}].
Grader 3 (YY) gave higher scores than the other graders, resulting in significantly improved results with hardware suppression ($p$ < 0.05 for NC vs H and S vs HS) and software correction ($p$ < 0.05 for H vs HS).
This higher visibility for this grader may be because this grader is a DOCT specialist and very familiar with LIV imaging, whereas the other graders, being skin specialists, are not highly familiar with LIV images.

The results of the sharpness evaluation (orange) were similar to those of the visibility assessment.
Namely, for the outer forearm, hardware correction notably improved the sharpness (NC vs H and S vs HS, $p$ < 0.01 for all graders).
However, the improvements were less pronounced for the inner forearm. 

Overall, these results demonstrate the significance of hardware motion suppression and the positive effect of additional software correction.

\subsubsection{Inter-grader agreement}
The inter-grader agreements of visibility and sharpness were assessed using the Spearman's correlation coefficients, which are summarized in Table S5. 
The  inter-grader agreements were found to be low, the mean $\pm$ standard deviations of the correlation coefficients were 0.62 $\pm$ 0.17 for the visibility and 0.57 $\pm$ 0.21 for the sharpness, which may be attributed to variations in graders' understanding of the verbally described visibility and sharpness criteria, especially between grades 1 and 2.

To verify this hypothesis, we reclassified the visibility scores into binary scores, where the cases with original scores of 1 or 2 were classified as ``visible,'' and those with an original score of 0 were classified as ``invisible.''
The agreements of binary scores between graders were computed by computing the agreement percentages, as shown in Table S6 in the supplementary material.
This analysis demonstrates high inter-grader agreement (with agreement more than 90\% for 40 out of 48 cases) and supports our hypothesis that the relatively low Spearman's correlation coefficients resulted from differences in understanding of the visibility criteria between scorers of 1 and 2.

Note that this analysis was not performed for the sharpness, because the results were, in principle, identical to those of visibility.
Specifically, the definition of a score of 0 for both visibility and sharpness was identical, defined as ``invisible.''
As a result, the binary scores for sharpness inherently become the analysis of ``invisible'' and ``visible,'' just like visibility.

\section{Discussion}
\subsection{Validation by \invitro skin model measurement}
To further understand the IEHV layer in the LIV images, we measured a reconstructed full thickness \invitro skin model (T-skin, Episkin, France).
T-skin features a fully stratified, differentiated, and self-renewing epidermis (including the four epidermal layers shown in Fig.\@ \ref{fig:Epidemic_layer}), along with a dermal compartment containing functional fibroblasts \cite{Bataillon_Tskin_2019}.

The T-skin at the proliferation stage was measured using DOCT with the same scanning protocol as for the \invivo human skin measurement but without hardware motion suppression and software motion correction.
After the DOCT measurement, the sample was processed for histological staining, including hematoxylin eosin (HE), filaggrin (FLG), and Ki67.

\begin{figure}
	\centering
	\includegraphics{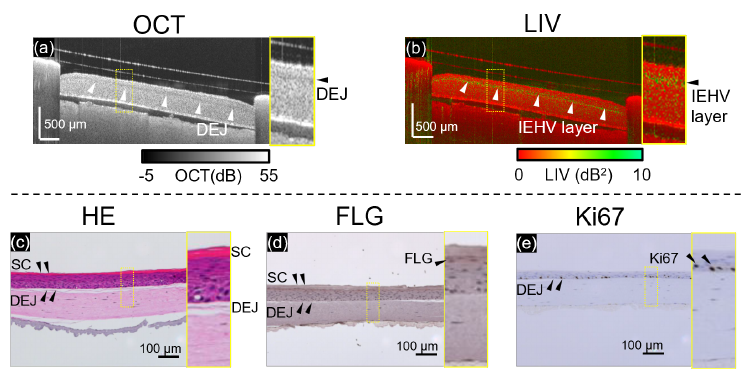}
	\caption{%
        (a) OCT, (b) LIV, and (c--e) histological micrographs of the skin organoid, T-skin, respectively.
        All images are vertical cross-sections.
        The IEHV layer appeared at the bottom of the epidermis in the LIV (b).
        The HE histology (c) exhibited distinctive contrasts of the stratum corneum (SC), epidermis, and dermis, which had distinctive colors.
        FLG in (d) is a differentiation marker that accumulates just beneath the SC.
        Ki67 (e) is the marker for proliferation that appears at the bottom of the epidermis.
        DEJ indicate the dermal-epidermal junction.
	}
	\label{fig:T-skin}
\end{figure}
The images are summarized in Fig.\@ \ref{fig:T-skin}.
The epidermis is known to exhibit relatively low scattering compared to the dermis \cite{Yamazaki_PSOCT_2020}, thereby enabling the identification of dermal-epidermal junction (DEJ) in the OCT image [white arrowheads in Fig.\@ \ref{fig:T-skin}(a)].
By comparing the OCT and DOCT images, we identified that the IEHV layer of the LIV [white arrowheads in Fig.\@ \ref{fig:T-skin}(b)] was located around the DEJ.
In the HE histology micrographs [Fig.\@ \ref{fig:T-skin}(c)], the stratum corneum (SC), epidermis, and dermis are distinctly visualized with red, dark purple, and pink colors, respectively, allowing the identification of the SC and DEJs.
In the FLG-staining micrograph [Fig.\@ \ref{fig:T-skin}(d)], FLG (brown color), which is a marker of cell differentiation, can be observed just beneath the SC.
By contrast, in the Ki67-staining micrograph [Fig.\@ \ref{fig:T-skin}(e)], Ki67 (dark brown), which is a marker of proliferation, appeared aound the DEJ.
The IEHV layer of the LIV appeared around the DEJ, and it seems collocating with the proliferation marker.

Similar to the T-skin case, the IEHV layer of the \invivo LIV [in Fig.\@ \ref{fig:IEHV_layer}] was located just around the DEJ.
This suggests that, in the \invivo cases, the IEHV layer may also indicate cellular proliferation.
Since keratinocyte proliferation is predominantly observed in the basal layer (i.e., SB) of the epidermis \cite{moreci_epidermal_2020}, the IEHV layer may indicate SB.

Because this discussion is based on a single sample, our interpretation of IEHV is only weakly supported. 
Further validation with additional samples and biopsy analysis are important to consolidate the conclusion.

\subsection{Blood-flow imaging}
\label{sec:bfImaging}
Although the primary target of current study was to visualize the intracellular motility of \invivo skin, DOCT was also found to be sensitive to tiny blood flow, as demonstrated in Section \ref{sec:resultSlabProjection}.
In general, it is because the flow sensitivity of DOCT and OCTA positively correlates with both the number of frames acquired at a single position and the time window.
In our case of DOCT imaging, 32 frames were acquired at each location with a time window as long as 6.55 s.

\begin{figure}
    \centering
    \includegraphics{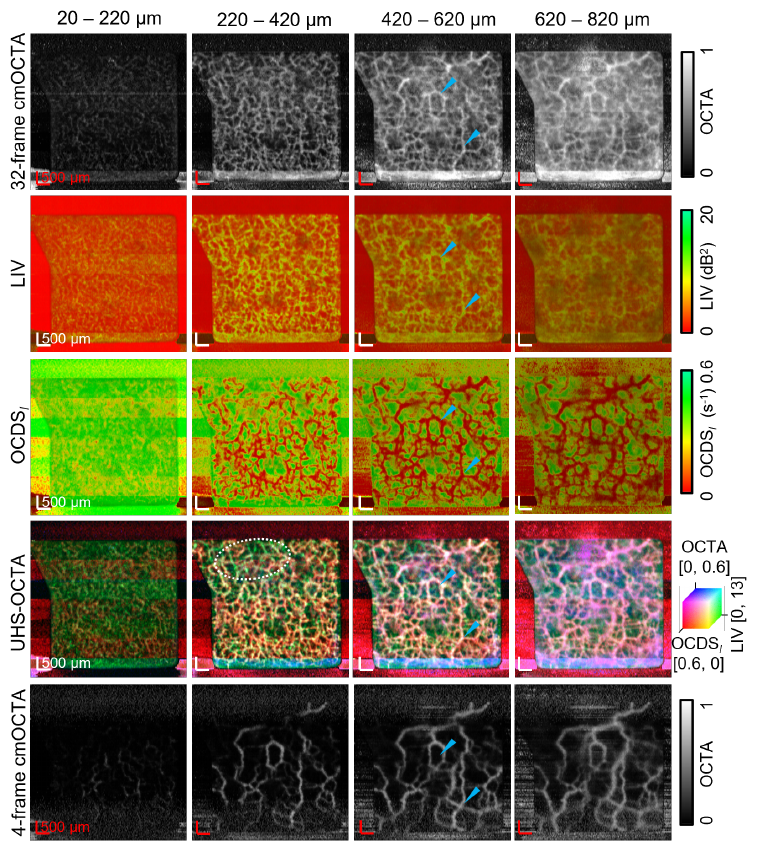}
    \caption{%
    Comparison of different vascular imaging methods across four depth ranges (20--220 \um, 220--420 \um, 420--620 \um, and 620--820 \um). 
    From the top, each row represents 32-frame cmOCTA, LIV, \OCDSl, and UHS-OCTA, and standard 4-frame cmOCTA.
    The UHS-OCTA was generated by integrating 32-frame cmOCTA, LIV, and \OCDSl.
    The scale bar represents 500 \um.
    The UHS-OCTA revealed significantly more small vessels than the standard 4-frame cmOCTA (bottom).
    } 
    \label{fig:UHS-OCTA}
\end{figure}

Here we demonstrate a preliminary method to generate ultra-high-sensitive (UHS-) OCTA from the dataset obtained using the 3D DOCT scanning protocol.
In this generation process, we first computed three types of DOCT and OCTA images, as shown in Fig.\@ \ref{fig:UHS-OCTA}.
One of them was the LIV, and another was the late OCT correlation decay speed (\OCDSl) \cite{el-sadek_optical_2020}, which is yet another DOCT method sensitive to the speed of the motion.
In addition, we computed the cmOCTA \cite{makita_noise-immune_2016} using 32 frames.
For this cmOCTA computation, the product of a frame and the complex conjugate of the adjacent frame were computed for each frame pair.
Then, the products of all pairs were averaged (Eq.\@ (30) of Ref.\@ \cite{makita_noise-immune_2016}).
This value was then normalized to within a range of 0 to 1 to obtain the correlation coefficient $\rho$ (Eq.\@ (31) of Ref.\@ \cite{makita_noise-immune_2016}).
The decorrelation (i.e., 1-$\rho$) was then used as the 32-frame cmOCTA.
In our specific implementation, we also applied a noise-effect correction, as detailed elsewhere \cite{makita_noise-immune_2016}.
Finally, the comprehensive UHS-OCTA was generated through the fusion of the LIV, \OCDSl, and 32-frame cmOCTA using a tri-variate color map.

Figure \ref{fig:UHS-OCTA} shows the DOCT and UHS-OCTA images of one subject's outer forearm at different depths. 
The standard 4-frame cmOCTA is also shown for reference, which was obtained from the same subject and at the same location but was acquired using the standard 4-frame repeated OCTA protocol.
The thick horizontal bars in \OCDSl and the UHS-OCTA are motion artifacts caused by the uncorrected sample motion at the particular block of the scan, so-called block artifacts.
We note that the \enface region is split into eight blocks in the 3D DOCT scanning protocol, as explained in Section \ref{sec:scanProtocol}.

Despite the block artifacts, in the superficial slab (20--220 \um), UHS-OCTA exhibited a very fine capillary network with high contrast, whereas this feature was dim or invisible in the other images.
In the deeper slabs (220--420 \um, 420--620 \um, and 620--820 \um), UHS-OCTA achieved remarkably high-contrast visualization of the vascular network.
Furthermore, the vessels became thicker in the deeper slabs, whereas the vascular network became denser at shallower depths.

\subsection{Why does simple motion suppression/correction enable \invivo DOCT?}
Since our motion-detection method detects motion with a 10-fold up-sampling factor (see Section \ref{sec:softCorrection}), the best achievable motion correction accuracy is 0.195 \um (lateral) and 0.724 \um (axial), which is one-tenth of the pixel separation.
This accuracy is a significant fraction of the wavelength.
However, the practically achieved motion-correction accuracy can be much lower than this maximum accuracy, potentially on the micrometer scale.
By contrast, intracellular motions are much smaller than the cell, typically involving only a small fraction of the cell's size.
This raises the question of how our relatively simple motion-suppression and motion-correction method enabled \invivo DOCT imaging.
This can be explained by categorizing the motions into bulk and intracellular motion.

The bulk motion of the sample is characterized by simultaneous shifts of all scatterers in the sample in the same direction, where the spatial relationships among scatterers are preserved.
As a result, the speckle pattern shifts uniformly without altering its structure.
The size of the speckle is approximately the size of the optical resolution, which was 18 \um (lateral) and 14 \um (axial) in our case.
Because this size was far larger than the motion-correction accuracy, the uncorrected (i.e., residual) motion did not significantly alter the OCT intensity at each point in the image.
Hence, the residual motion did not cause remarkable motion artifacts, such as an elevation of DOCT values.

On the other hand, the relative positions and distances of the scatterers were randomized by the intracellular motion.
In this case, even the displacements of the scatterers were far shorter than the wavelength, which caused significant random alterations in the interference pattern (i.e., the shape of speckles).
Therefore, DOCT is more sensitive to the intracellular motion than the bulk motion.

Specifically, by suppressing and correcting the bulk motion to within around a micrometer, the DOCT becomes insensitive to the bulk motion and can highlight the intracellular motions.

It is also noteworthy that a similar discussion can be applicable to OCTA.
Note that, for vascular imaging, some cells in resolution volume, such as red blood cells, travel in the same direction and at the same speed, whereas others travel with different speeds.
Furthermore, some tissues, such as the vessel wall, remain static.
Hence, the speckle pattern not only shifts but also alters by the flow, whereas bulk motion only shifts the speckle pattern.

\subsection{Future perspectives}
One of the challenges of hardware motion suppression is the trade-off between the motion-reduction capability and the preservation of physiological conditions. 
Applying a higher pressure with the fixation attachment reduces more bulk motion but may occlude the blood flow \cite{shi_integrating_2022} and alter other physiological conditions.
A possible solution is the integration of a pressure sensor to the fixation attachment, similar to that demonstrated in \cite{shi_integrating_2022}.
This may allow for the precise control of the applied pressure and enable a balance to be achieved between tissue stabilization and the preservation of physiological conditions. 

Neural network (NN) may further improve \invivo DOCT.
For example, a NN-based motion-artifact correction was demonstrated in OCTA \cite{li_deep-learning-based_2021, Lin_2024_DeeplearnOCTA}.
Similarly, an NN-based method could be applied for the motion-artifact correction of DOCT.

Liu \etal demonstrated NN-based fast LIV imaging, which uses only four frames at a given location, and hence achieves an eightfold reduction of volumetric measurement time \cite{liu_neural-network_2024}.
In our protocol, this method would result in a volumetric measurement time of less than 7 s and would significantly enhance subject comfort and facilitate \invivo DOCT measurements.

\section{Conclusions}
Hardware motion suppression (via fixation attachment) and image-correlation-based software motion correction were applied to \invivo DOCT imaging.
Subjective image observation, quantitative evaluation of motion artifacts through ROI-based mean LIV analysis, and grading-based qualitative assessment of the dynamic structure visibility supported the effectiveness of our relatively simple motion-suppression and motion-correction method.
In conclusion, the presented method enables \invivo DOCT imaging of human skin.

\section*{Funding}
Core Research for Evolutional Science and Technology (JPMJCR2105); 
Japan Society for the Promotion of Science (21H01836, 22F22355, 22KF0058, 22K04962, 24KJ0510);
University of Tsukuba (Tsukuba Scholarship);
Japan Science and Technology Agency (JPMJFS2106). 

\section*{Disclosures}
Guo, Morishita, El-Sadek, Mukherjee, Lim, Bao, Makita, Yasuno: Sky Technology(F), Nikon(F), Kao Corp.(F), Topcon(F), Panasonic(F), Santec (F), Nidek (F).
Yamazaki, Sakai, Sugata, Kasamatsu, Yoshida: Kao Corp. (E).
Guo and Lim are currently employed by Ainnovi.
	
\section* {Data, Materials, and Code Availability} 
Data underlying the results presented in this paper are not publicly available at this time but may be obtained from the authors upon reasonable request. 
The CAD design of the fixation attachment and code of the software motion correction can be found in the following online repository.
\url{https://github.com/ComputationalOpticsGroup/SkinDOCT-motion-correction/} \cite{GitHubMotionCorrection}

\section*{Supplemental document}
See Supplement 1 for supporting content.

\bibliography{reference}

\newpage


\pagebreak
\title{Supplementary Material}
\setcounter{figure}{0}
\renewcommand\thefigure{S\arabic{figure}}   
\setcounter{table}{0}
\renewcommand\thetable{S\arabic{table}}   

\setcounter{section}{0}
\renewcommand\thesection{S\arabic{section}}

\begin{abstract}
	In this supplementary material, we summarize the results of statistical analyses across six tables, including $p$-values from t-tests and Wilcoxon signed-rank tests, as well as inter-grader agreements.
	Tables \ref{tab:S1_p_values_outer_forearm} and \ref{tab:S2_p_values_inner_forearm} supplement Section 3.2 and Fig.\@ 7 of the main manuscript.
	Table \ref{tab:S1_p_values_outer_forearm} displays the $p$-values from the t-test comparing various motion suppression/correction configurations at the outer forearm across three regions of interest (ROIs). 
	Similarly, Table \ref{tab:S2_p_values_inner_forearm} shows the p-values at the inner forearm.
	The results of the Wilcoxon signed-rank test for the visibility and sharpness evaluation of the intra-epidermal high-value (IEHV) layer are presented in Table \ref{tab:S3_p_visibility} and Table \ref{tab:S4_p_sharpness}, respectively.
	These two tables supplement Section 3.3 and Fig.\@ 9 of the main manuscript.
	Additionally, the inter-grader agreements for visibility and sharpness were evaluated using Spearman's correlation coefficients, which are summarized in Table \ref{tab:S5_correlation}. 
	The agreement percentages for binary-visibility scores between graders are presented in Table \ref{tab:S6_binary_visibility}.
	These two tables supplement Section 3.3.2 of the main manuscript.
\end{abstract}


\begin{table}
	\centering
	\caption{
		The t-test results comparing different motion suppression/correction configurations at the outer forearm across three ROIs.
		Each cell displays the $p$-value for the comparison between two configurations. 
	}
	\label{tab:S1_p_values_outer_forearm}
	\setlength{\tabcolsep}{6pt}
	\adjustbox{max width=\textwidth}{
		\begin{tabular}{@{}lcccc|cccc|cccc@{}}
			\toprule
			\textbf{Outer-forearm} & \multicolumn{4}{c}{\textbf{ROI 1}} & \multicolumn{4}{c}{\textbf{ROI 2}} & \multicolumn{4}{c}{\textbf{ROI 3}} \\
			\cmidrule(lr){2-5} \cmidrule(lr){6-9} \cmidrule(lr){10-13}
			& \textbf{NC} & \textbf{S} & \textbf{H} & \textbf{HS} 
			& \textbf{NC} & \textbf{S} & \textbf{H} & \textbf{HS} 
			& \textbf{NC} & \textbf{S} & \textbf{H} & \textbf{HS} \\ 
			\midrule
			\textbf{NC} & -- & 0.000147 & 0.000009 & 0.000006 
			& -- & 0.000025 & 0.000000 & 0.000000 
			& -- & 0.000031 & 0.000000 & 0.000000 \\
			\textbf{S} & -- & -- & 0.000004 & 0.000121 
			& -- & -- & 0.000004 & 0.000089 
			& -- & -- & 0.000025 & 0.000070 \\
			\textbf{H} & -- & -- & -- & 0.000004 
			& -- & -- & -- & 0.000004 
			& -- & -- & -- & 0.000004 \\
			\textbf{HS} & -- & -- & -- & -- 
			& -- & -- & -- & -- 
			& -- & -- & -- & -- \\
			\bottomrule
		\end{tabular}
	}
\end{table}

\begin{table}
	\centering
	\caption{
		The t-test results comparing different motion suppression/correction configurations at the inner forearm across three ROIs.
		Each cell displays the $p$-value for the comparison between two configurations.  
	}
	\label{tab:S2_p_values_inner_forearm}
	\setlength{\tabcolsep}{6pt}
	\adjustbox{max width=\textwidth}{
		\begin{tabular}{@{}lcccc|cccc|cccc@{}}
			\toprule
			\textbf{Inner-forearm} & \multicolumn{4}{c}{\textbf{ROI 1}} & \multicolumn{4}{c}{\textbf{ROI 2}} & \multicolumn{4}{c}{\textbf{ROI 3}} \\
			\cmidrule(lr){2-5} \cmidrule(lr){6-9} \cmidrule(lr){10-13}
			& \textbf{NC} & \textbf{S} & \textbf{H} & \textbf{HS} 
			& \textbf{NC} & \textbf{S} & \textbf{H} & \textbf{HS} 
			& \textbf{NC} & \textbf{S} & \textbf{H} & \textbf{HS} \\ 
			\midrule
			\textbf{NC} & -- & 0.001404 & 0.000076 & 0.000052 
			& -- & 0.000000 & 0.000000 & 0.000000 
			& -- & 0.000001 & 0.000000 & 0.000000 \\
			\textbf{S} & -- & -- & 0.000000 & 0.007712 
			& -- & -- & 0.000000 & 0.005341 
			& -- & -- & 0.000002 & 0.003609 \\
			\textbf{H} & -- & -- & -- & 0.000000 
			& -- & -- & -- & 0.000000 
			& -- & -- & -- & 0.000000 \\
			\textbf{HS} & -- & -- & -- & -- 
			& -- & -- & -- & -- 
			& -- & -- & -- & -- \\
			\bottomrule
		\end{tabular}
	}
\end{table}

\begin{table}
	\centering
	\caption{
		The Wilcoxon signed-rank test results of visibility analysis for the IEHV layer.
		Columns 3--10 (NC, S, H, HS, NC, S, H, and HS) show the $p$-values for pairwise comparisons of different motion suppression/correction configurations within the same imaging method (OCT or LIV).
		Columns 11--12 (Outer and Inner) show the $p$-values for comparisons between OCT and LIV imaging methods under identical configurations.
		NA indicates cases where the Wilcoxon signed-rank test could not be performed due to all scores from one or both graders being zero.
	}
	\label{tab:S3_p_visibility}
	\renewcommand{\arraystretch}{1.2}
	\adjustbox{max width=\textwidth}{
		\begin{tabular}{@{}clcccc|cccc|cc@{}}
			\toprule
			\multicolumn{2}{c}{} & \multicolumn{4}{c}{\textbf{LIV (Outer)}} & \multicolumn{4}{c}{\textbf{LIV (Inner)}} & \multicolumn{2}{c}{\textbf{LIV vs OCT}} \\ 
			\cmidrule(lr){3-6} \cmidrule(lr){7-10} \cmidrule(lr){11-12}
			\textbf{Grader} & \textbf{Configure} & NC & S & H & HS & NC & S & H & HS & Outer & Inner \\ 
			\midrule
			\multirow{4}{*}{SS} 
			& NC & -- & NA & 0.0094 & 0.0062 & -- & NA & 0.0588 & 0.1025 & NA & NA \\ 
			& S & -- & -- & 0.0094 & 0.0062 & -- & -- & 0.0588 & 0.1025 & NA & NA \\ 
			& H & -- & -- & -- & 0.1573 & -- & -- & -- & 0.1025 & 0.0094 & 0.0588 \\ 
			& HS & -- & -- & -- & -- & -- & -- & -- & -- & 0.0062 & 0.1025 \\ 
			\midrule
			\multirow{4}{*}{KY}
			& NC & -- & NA & 0.0231 & 0.0094 & -- & NA & 0.0833 & 0.1573 & NA & NA \\ 
			& S & -- & -- & 0.0231 & 0.0094 & -- & -- & 0.0833 & 0.1573 & NA & NA \\ 
			& H & -- & -- & -- & 0.0833 & -- & -- & -- & 0.3173 & 0.0038 & 0.1573 \\ 
			& HS & -- & -- & -- & -- & -- & -- & -- & -- & 0.0094 & 0.1573 \\ 
			\midrule
			\multirow{4}{*}{YY} 
			& NC & -- & NA & 0.0062 & 0.0020 & -- & NA & 0.0384 & 0.0196 & NA & NA \\ 
			& S & -- & -- & 0.0062 & 0.0020 & -- & -- & 0.0384 & 0.0196 & NA & NA \\ 
			& H & -- & -- & -- & 0.1797 & -- & -- & -- & 0.0384 & 0.0062 & 0.0384 \\ 
			& HS & -- & -- & -- & -- & -- & -- & -- & -- & 0.0020 & 0.0141 \\ 
			\bottomrule
		\end{tabular}
	}
\end{table}

\begin{table}
	\centering
	\caption{
		The Wilcoxon signed-rank test results of sharpness analysis for the IEHV layer.
		Columns 3--10 (NC, S, H, HS, NC, S, H, and HS) show the $p$-values for pairwise comparisons of different motion suppression/correction configurations within the same imaging method (OCT or LIV).
		Columns 11--12 (Outer and Inner) show the $p$-values for comparisons between OCT and LIV imaging methods under identical configurations.
		NA indicates cases where the Wilcoxon signed-rank test could not be performed due to all scores from one or both graders being zero.
	}
	\label{tab:S4_p_sharpness}
	\renewcommand{\arraystretch}{1.2}
	\adjustbox{max width=\textwidth}{
		\begin{tabular}{@{}clcccc|cccc|cc@{}}
			\toprule
			\multicolumn{2}{c}{} & \multicolumn{4}{c}{\textbf{LIV (Outer)}} & \multicolumn{4}{c}{\textbf{LIV (Inner)}} & \multicolumn{2}{c}{\textbf{LIV vs OCT}} \\ 
			\cmidrule(lr){3-6} \cmidrule(lr){7-10} \cmidrule(lr){11-12}
			\textbf{Grader} & \textbf{Configure} & NC & S & H & HS & NC & S & H & HS & Outer & Inner \\ 
			\midrule
			\multirow{4}{*}{SS} 
			& NC & -- & NA & 0.0097 & 0.0058 & -- & NA & 0.0588 & 0.1025 & NA & NA \\ 
			& S & -- & -- & 0.0097 & 0.0058 & -- & -- & 0.0588 & 0.1025 & NA & NA \\ 
			& H & -- & -- & -- & 0.0833 & -- & -- & -- & 0.1797 & 0.0097 & 0.0588 \\ 
			& HS & -- & -- & -- & -- & -- & -- & -- & -- & 0.0058 & 0.1025 \\ 
			\midrule
			\multirow{4}{*}{KY}
			& NC & -- & NA & 0.0139 & 0.0083 & -- & NA & 0.1573 & 0.1797 & 0.1573 & NA \\ 
			& S & -- & -- & 0.0139 & 0.0083 & -- & -- & 0.1573 & 0.1797 & 0.3173 & NA \\ 
			& H & -- & -- & -- & 0.0412 & -- & -- & -- & 0.3173 & 0.0633 & 0.3173 \\ 
			& HS & -- & -- & -- & -- & -- & -- & -- & -- & 0.0083 & 0.1797 \\ 
			\midrule
			\multirow{4}{*}{YY} 
			& NC & -- & NA & 0.0058 & 0.0020 & -- & NA & 0.0339 & 0.0244 & NA & NA \\ 
			& S & -- & -- & 0.0058 & 0.0020 & -- & -- & 0.0339 & 0.0244 & NA & NA \\ 
			& H & -- & -- & -- & 0.0833 & -- & -- & -- & 0.0656 & 0.0058 & 0.0339 \\ 
			& HS & -- & -- & -- & -- & -- & -- & -- & -- & 0.0020 & 0.0164 \\ 
			\bottomrule
		\end{tabular}
	}
\end{table}

\begin{table}
	\centering
	\caption{%
		Inter-grader agreement of visibility and sharpness. 
		Each cell shows the Spearman's correlation coefficient.
		NA stands for not available, indicating cases in which all scores of one or both graders were zero, such that the correlation coefficient could not be computed.
	}
	\label{tab:S5_correlation}
	\begin{tabular}{@{}c c cccc cccc@{}}
		\toprule
		\textbf{} & \textbf{Graders} & \multicolumn{4}{c}{\textbf{LIV at outer-forearm}} & \multicolumn{4}{c}{\textbf{LIV at inner-forearm}} \\ 
		\cmidrule(lr){3-6} \cmidrule(lr){7-10}
		& & \textbf{NC} & \textbf{S} & \textbf{H} & \textbf{HS} & \textbf{NC} & \textbf{S} & \textbf{H} & \textbf{HS} \\ 
		\midrule
		\multirow{3}{*}{\centering Visibility} 
		& \centering SS-KY & NA & NA & 0.63 & 0.83 & NA & NA & 0.72 & 0.67 \\ 
		& \centering SS-YY & NA & NA & 0.57 & 0.35 & NA & NA & 0.83 & 0.81 \\ 
		& \centering KY-YY & NA & NA & 0.32 & 0.42 & NA & NA & 0.66 & 0.58 \\ 
		\midrule
		\multirow{3}{*}{\centering Sharpness} 
		& \centering SS-KY & NA & NA & 0.80 & 0.27 & NA & NA & 0.70 & 0.86 \\ 
		& \centering SS-YY & NA & NA & 0.54 & 0.44 & NA & NA & 0.79 & 0.36 \\ 
		& \centering KY-YY & NA & NA & 0.76 & 0.32 & NA & NA & 0.63 & 0.39 \\ 
		\bottomrule
	\end{tabular}
\end{table}

\begin{table}
	\centering
	\caption{%
		Agreement in the binarized visibility score, with each cell indicating the percentage of cases where two graders agreed. 
		These scores must be identical to the binarized sharpness score by definition (see text).
	}
	\label{tab:S6_binary_visibility}
	\renewcommand{\arraystretch}{1.2}
	\adjustbox{max width=\textwidth}{
		\begin{tabular}{@{}cccccccccc@{}}
			\toprule
			\textbf{Image type} & \textbf{Graders} & \multicolumn{4}{c}{\textbf{Outer-forearm}} & \multicolumn{4}{c}{\textbf{Inner-forearm}} \\ 
			\cmidrule(lr){3-6} \cmidrule(lr){7-10}
			& & \textbf{NC} & \textbf{S} & \textbf{H} & \textbf{HS} & \textbf{NC} & \textbf{S} & \textbf{H} & \textbf{HS} \\ 
			\midrule
			\multirow{3}{*}{LIV} 
			& SS-KY & 100\% & 100\% & 80\% & 90\% & 100\% & 100\% & 90\% & 90\% \\ 
			& SS-YY & 100\% & 100\% & 90\% & 90\% & 100\% & 100\% & 90\% & 70\% \\ 
			& KY-YY & 100\% & 100\% & 70\% & 80\% & 100\% & 100\% & 80\% & 60\% \\ 
			\midrule
			\multirow{3}{*}{OCT} 
			& SS-KY & 80\% & 90\% & 90\% & 100\% & 100\% & 100\% & 90\% & 100\% \\ 
			& SS-YY & 90\% & 100\% & 90\% & 100\% & 100\% & 100\% & 90\% & 100\% \\ 
			& KY-YY & 80\% & 90\% & 90\% & 100\% & 100\% & 100\% & 90\% & 100\% \\ 
			\bottomrule
		\end{tabular}
	}
\end{table}

\end{document}